\documentclass[aps, preprint, doi=false,prl,amsmath, citeautoscript]{revtex4-1}
\usepackage{graphics}
\usepackage{natbib}
\usepackage{graphicx}
\usepackage{epsfig}
\usepackage{amsmath}
\usepackage{bm}
\usepackage{lineno}
\usepackage{color}

\begin{document}
\title{Suppression of the fieldlike spin-orbit torque efficiency due to the magnetic proximity effect in ferromagnet/platinum bilayers}

\author{T. A. Peterson$^1$, A. P. McFadden$^2$, C. J. Palmstr{\o}m$^2$, and P. A. Crowell$^1$}
\date{\today}
\affiliation{$^1$School of Physics and Astronomy, University of Minnesota, Minneapolis, Minnesota 55455\\
$^2$Departments of Electrical \& Computer Engineering and
Materials, University of California, Santa Barbara, California
93106}
\begin{abstract}
Current-induced spin-orbit torques in Co$_2$FeAl/Pt ultrathin bilayers are studied using a magnetoresistive harmonic response technique, which distinguishes the dampinglike and fieldlike contributions. The presence of a temperature-dependent magnetic proximity effect is observed through the anomalous Hall and anisotropic magnetoresistances, which are enhanced at low temperatures for thin platinum thicknesses. The fieldlike torque efficiency decreases steadily as the temperature is lowered for all Pt thicknesses studied, which we propose is related to the influence of the magnetic proximity effect on the fieldlike torque mechanism.
\end{abstract}
\maketitle

Through the spin-orbit interaction (SOI), an electrical current $\bm{j}_e$ in a ferromagnet(F)/nonmagnetic metal(N) bilayer results in a torque on the magnetization $\bm{M}$ of F\cite{Miron2011,Liu2012}. This spin-orbit torque (SOT) may be decomposed into two perpendicular components -- a component oriented along $\hat{\bm{m}}\times(\hat{\bm{m}}\times\bm{\hat{\sigma}})$ and a component along $\hat{\bm{m}}\times\bm{\hat{\sigma}}$, where $\bm{\hat{\sigma}} \equiv \bm{\hat{j}}_e\times \hat{\bm{n}}$ denotes the orientation of the spin current created by the SOI and $\hat{\bm{n}}$ defines the unit vector normal to the plane formed by the F/N interface. These are referred to respectively as the dampinglike (DL) and fieldlike (FL) SOTs. Although the microscopic origins of the DL and FL SOTs remain unclear, the DL contribution has been widely interpreted using N bulk spin-Hall effect (SHE) diffusion models\cite{Liu2012,Liu_Pt_2012,Fan_2013,Nguyen2016}, and the FL contribution attributed to the F/N interfacial SOI\cite{Miron2011,Fan_2013}. \citet{Amin_Stiles_2016} have recently emphasized that this interpretation is overly simplistic, showing that the interfacial SOI and the SHE in the N layer may both produce FL \textit{and} DL torques depending on the interface details. Unfortunately, the interfacial parameters used in spin diffusion models are not easily measured, and it remains an experimental challenge to separately identify the origins of the DL and FL torques. Also, in bilayers where interface scattering is dominant, a conventional normal-to-interface spin diffusion length becomes difficult to define. Furthermore, magnetic proximity effects (MPE) at F/N interfaces have been widely reported\cite{Cheng2004,Huang2012,Zhang2015,Yang2014}, yet how the MPE influences SOTs is unknown.

In this article, we report a decrease in the FL and DL torques per unit current density (hereafter referred to as SOT efficiencies) at low temperature in Co$_2$FeAl/Pt bilayers. In the same bilayers, a temperature-dependent MPE is revealed through magnetoresistance (MR) measurements. The FL SOT efficiency is suppressed by nearly a factor of 4 at 20~K with respect to room temperature for all Pt thicknesses studied, which we propose is related to the increasing influence of the MPE exchange field on the F/N interface Rashba spin accumulation. Meanwhile, the DL SOT efficiency monotonically increases with decreasing Pt thickness and closely tracks the Pt resistivity as temperature is varied. Within the Pt SHE diffusion model, the latter observation may be described by either the intrinsic SHE or spin backflow processes, between which we cannot differentiate.

The F/N bilayers used in this study were grown on MgO(001) substrates by molecular-beam epitaxy~(MBE). Prior to F growth, an \textit{in-situ} MgO buffer was grown by e-beam evaporation on prepared MgO substrates in order to bury residual carbon and improve surface morphology. The F layer is the Heusler compound  Co$_2$FeAl~(CFA) with thickness $t_F = 1.2$~nm, grown by MBE at a substrate temperature of 200$^\circ$~C by codeposition of individual elemental sources in ultrahigh vacuum~(UHV). Reflection high energy electron diffraction~(RHEED) monitored during CFA growth confirmed a 45$^\circ$ rotated orientation CFA$<$110$>$~$\vert\vert$~MgO$<$100$>$. X-ray diffraction~(XRD) measurements conducted on thicker 4 and 30 nm MgO/CFA samples confirm a single phase of (001) oriented CFA while the presence and relative peak area of the (002) reflection confirms at least B2 ordering. The samples were cooled to room temperature before capping with Pt, which was grown using e-beam evaporation in UHV. The Pt grew epitaxially and was (001) oriented with Pt$<$100$>$~$\vert\vert$~CFA$<$110$>$, as confirmed by RHEED and XRD. An \textit{in-situ} shadowmask technique was used to achieve four different Pt cap thicknesses ($t_N$) on the same MgO/CFA(1.2~nm) underlayer. Two growths, one with $t_{N} = 1,2,3,4$~nm and the other with $t_{N} = 5,6,7,8$~nm, were used in this study. After Pt capping, samples were removed from UHV and exposed to atmosphere for subsequent processing. Vibrating sample magnetometry was used to measure the CFA(1.2~nm) saturation magnetization $M_s = 800\pm100$~emu/cm$^3$ at room temperature. The saturation magnetic field of the anomalous Hall effect (AHE) at 300~K matched $4\pi M_s$ within uncertainty. Therefore, the AHE saturation field was used to infer the temperature dependence of $M_s$, which increased from $850$~emu/cm$^3$ at 300~K to $1050$~emu/cm$^3$ at 10 K. Ferromagnetic resonance (FMR) measurements were performed at room temperature on a companion MgO/CFA(1.2~nm)/Pt(7~nm) bilayer, for which Kittel formula\cite{Kittel1948} fits of the FMR field for rf excitation frequencies from 4-20 GHz revealed a cubic in-plane anisotropy $K_1=-6\times 10^3$~J/m$^3$ with magnetic easy axes along CFA$<$110$>$(MgO$<$100$>$).

The bilayers were patterned into Hall bars by photolithography and Ar$^+$-ion milling, and Ti/Au vias and bonding pads were subsequently deposited. The Hall bar width was 10~$\mu$m. A magnetoresistive second harmonic (2$\omega$) response technique similar to that discussed in Refs. \cite{Fan_2013, Kawaguchi2013,Avci2014} was employed to measure the SOT efficiencies. The DL and FL effective fields $H_{DL}$ and $H_{FL}$ result in 2$\omega$ Hall resistances due to the anomalous Hall effect (AHE) and planar Hall effect (PHE), respectively. An applied magnetic field was rotated 360$^\circ$ in the sample plane, and the angular dependence of the 2$\omega$ Hall resistance was fit to extract $H_{DL}$ and $H_{FL}$. Magnetothermoelectric effects\cite{Bauer_2012}, which can contribute to 2$\omega$ resistances, were carefully taken into account. See the Supplemental Material\cite{CFA_Pt_supplementary} for a detailed description of the measurement geometry and fitting procedure. The dimensionless SOT efficiency is given by\cite{Pai2015}
\begin{equation}\label{xi}
\xi_{DL(FL)} \equiv \frac{M_s t_{F} H_{DL(FL)}}{(\hbar/2e)j^e_N},
\end{equation}
where $e$ is the electron charge, $\hbar$ is Planck's constant, and $j^e_N$ is the current density in the N layer.
\begin{figure}
\includegraphics{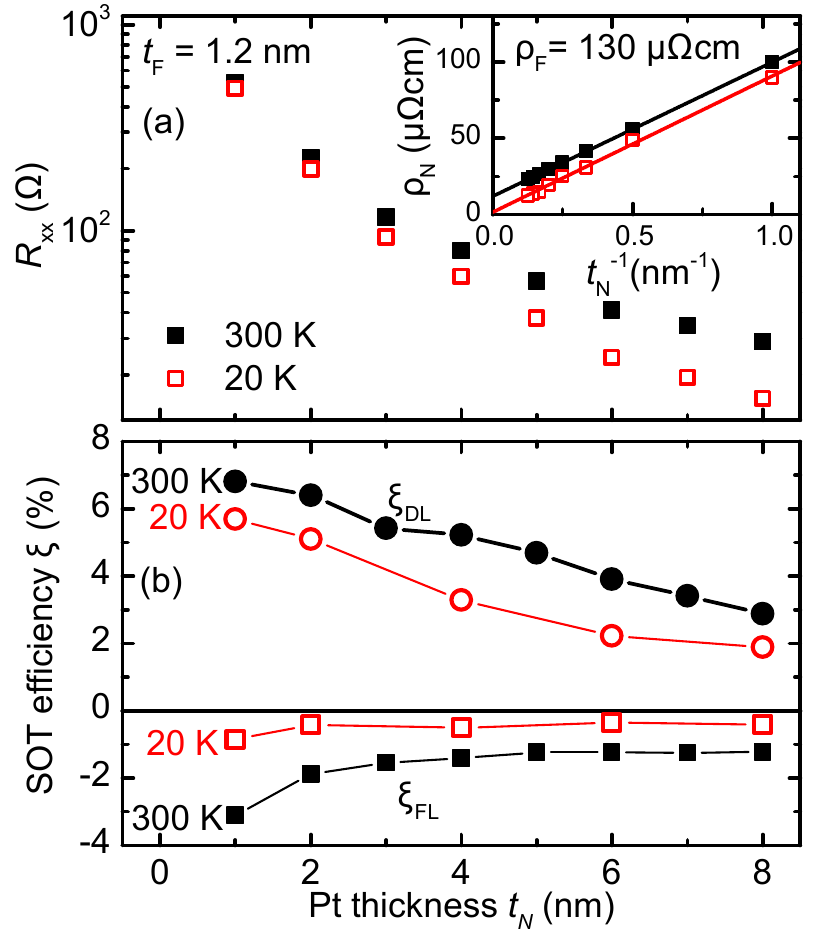}
\caption{(a) The bilayer square resistance for all Pt thicknesses. The solid black squares are 300~K data and open red squares are 20~K data. In the inset, the Pt resistivity is plotted vs. the inverse of the Pt thickness. The intercepts of the solid lines correspond to the bulk resistivity of Pt. In (b), the SOT efficiencies $\xi_{DL}$ (circles) and $\xi_{FL}$ (squares) are shown for different Pt thicknesses at 300~K (black solid symbols) and 20~K (red open symbols). The lines connect data points. For all data the CFA thickness is 1.2~nm.} \label{fig:RT_summ}
\end{figure}

\begin{figure}
\includegraphics{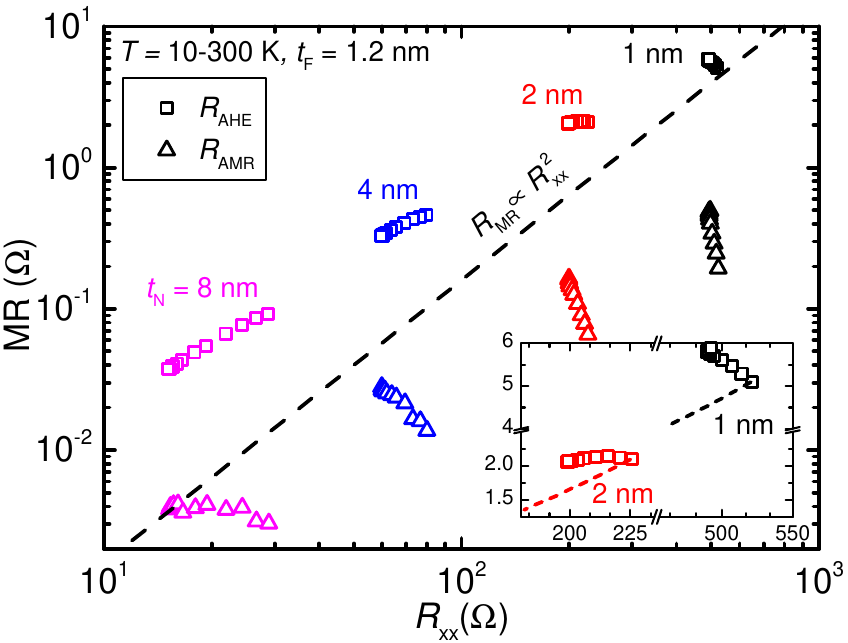}
\caption{Summary of $R_{AHE}$ (squares) and $R_{AMR}$ (triangles) vs. $R_{xx}$ for the different bilayers which are labeled by the Pt thickness. Temperature is an implicit variable, and the minima and maxima of $R_{xx}$ correspond to 10 and 300~K, respectively, for all bilayers except the 1~nm Pt bilayer, in which $R_{xx}$ shows a small upturn below 20~K. The dashed lines indicate $R_{MR} \propto R_{xx}^2$, which is expected for MR originating from F shunting alone. The inset magnifies the AHE data for the 1 and 2~nm Pt bilayers. See Fig. 4 in the Supplemental Material\cite{CFA_Pt_supplementary} for an alternative representation in which temperature is indicated explicitly, and details on how $R_{AHE}$ and $R_{AMR}$ were measured. } \label{fig:Rxx_Rxy}
\end{figure}

The bilayer square resistances $R_{xx}$ are summarized in Fig. \ref{fig:RT_summ}(a) for all Pt thicknesses at temperatures of 300~K and 20~K. The inset of Fig. \ref{fig:RT_summ}(a) shows the Pt resistivity, which is a strong function of thickness due to diffuse surface scattering\cite{Sonheimer_1952}. The F and N layers are treated as parallel resistances to account for the current shunted through F and determine $j^e_N$ in the denominator of Eq. \ref{xi}. See the Supplemental Material\cite{CFA_Pt_supplementary} for a detailed discussion of the shunting model and the method used to extract the Pt and CFA resistivities from $R_{xx}$. The CFA resistivity extracted from the shunting model is 130~$\mu\Omega$cm, which is similar to resistivities we measure for thicker 5 and 10~nm CFA films capped with AlOx. For the 5 and 10~nm CFA films, resistivities are $\rho\simeq100~\mu\Omega$cm with residual resistivity ratios RRR$\simeq 1.1$, and we have also measured the AHE resistivity $\rho_{AHE}\simeq0.6~\mu\Omega$cm. For these CFA films we find $\rho_{AHE}$ decreases as temperature is decreased, with a trend close to $\rho_{AHE} \propto \rho^2$. In contrast, for the CFA(1.2~nm)/Pt bilayers we observe an increase in the AHE resistance $R_{AHE}$ and anisotropic MR $R_{AMR}$ at low temperatures for thin Pt thicknesses. ($R_{AHE}$ is defined by the expression $R_{xy}=R_{AHE}m_z+R_H$, with $m_z$ denoting the out-of-plane magnetization component and $R_H$ the ordinary Hall effect resistance, and $R_{AMR}\equiv (R^{\vert\vert}_{xx}-R^{\perp}_{xx})/2$ with the parallel and perpendicular superscripts denoting the orientation of the current and saturated magnetization.) Figure \ref{fig:Rxx_Rxy} summarizes the temperature and Pt thickness dependence of $R_{AHE}$ and $R_{AMR}$ by plotting these MRs vs. $R_{xx}$, in which temperature is the implicit variable. The temperature was varied between 10~K (low $R_{xx}$) and 300~K (high $R_{xx}$). (See the Supplemental Material\cite{CFA_Pt_supplementary} for example magnetic field sweeps used to extract $R_{AHE}$ and $R_{AMR}$, and for an alternative representation of the data shown in Fig. \ref{fig:Rxx_Rxy} in which temperature is indicated explicitly.)

The increase in the (extraordinary, or anomalous\cite{Mcguire1975}) MR observed at low temperatures in Fig. \ref{fig:Rxx_Rxy} is due to the MPE. Because of current shunting through the F in metallic F/N bilayers, MR-based studies of the MPE have typically been relegated to ferromagnet insulator/Pt bilayers\cite{Huang2012,Lu2013,Miao2014,Miao2017}. However, the MR behavior shown in Fig. \ref{fig:Rxx_Rxy} as temperature is decreased cannot be attributed to shunting through F. Given F RRR values near unity, F shunting alone results in a measured $R_{MR}\propto R_{xx}^2$\cite{CFA_Pt_supplementary}. In fact, the trends of both $R_{AHE}$ and $R_{AMR}$ consistently show excess MR at low temperature compared to the $R_{MR}\propto R_{xx}^2$ trend drawn on Fig. \ref{fig:Rxx_Rxy}, indicating an additional MPE MR contribution at low temperature. Furthermore, for the 1 and 2~nm Pt bilayers, both AHE and AMR resistances \textit{increase} as the temperature decreases. For the 1~nm Pt bilayer $R_{AMR}$ increases by a factor of 3 from 300~K to 10~K, in stark contrast to the F shunting prediction of a 12\% decrease over the same temperature range. In fact, the bilayer $R_{AMR}>0$ is opposite in sign to that measured on 5~nm CFA films with Al capping layers, highlighting the influence of the Pt layer on the AMR.

Briefly, we discuss the relevance of the recently-discovered spin-Hall MR (SMR) effect\cite{Nakayama2013, Althammer2013, Meyer2014, Kim2016} to our MR measurements. The conventional AMR effect\cite{Mcguire1975} magnitudes summarized in Fig. \ref{fig:Rxx_Rxy} were obtained by performing the measurement in a geometry such that the SMR effect is absent, similar to Ref. \cite{Nakayama2013}. See the Supplemental Material\cite{CFA_Pt_supplementary} for the details of the measurement geometry used to differentiate $R_{AMR}$ from SMR effects. (We do observe a SMR-like MR of magnitude $\Delta R_{xx}/R_{xx}\sim10^{-3}$, but these effects are not the focus of this letter.) It has been reported that the SMR effect in N may give rise to an AHE-like transverse resistance (SH-AHE)\cite{Althammer2013,Meyer2015,Chen2013}. In comparison to Refs. \cite{Althammer2013,Meyer2015}, however, in our bilayers $R_{AHE}$ is a factor of 10-100 times larger. Furthermore, given that we observe SMR magnitudes $\sim10^{-3}$, we expect the SH-AHE magnitude ($R_{xy}/R_{xx}$) to be of order $10^{-4}-10^{-5}$\cite{Chen2013}, much smaller than the AHE we observe.

\begin{figure}
\includegraphics{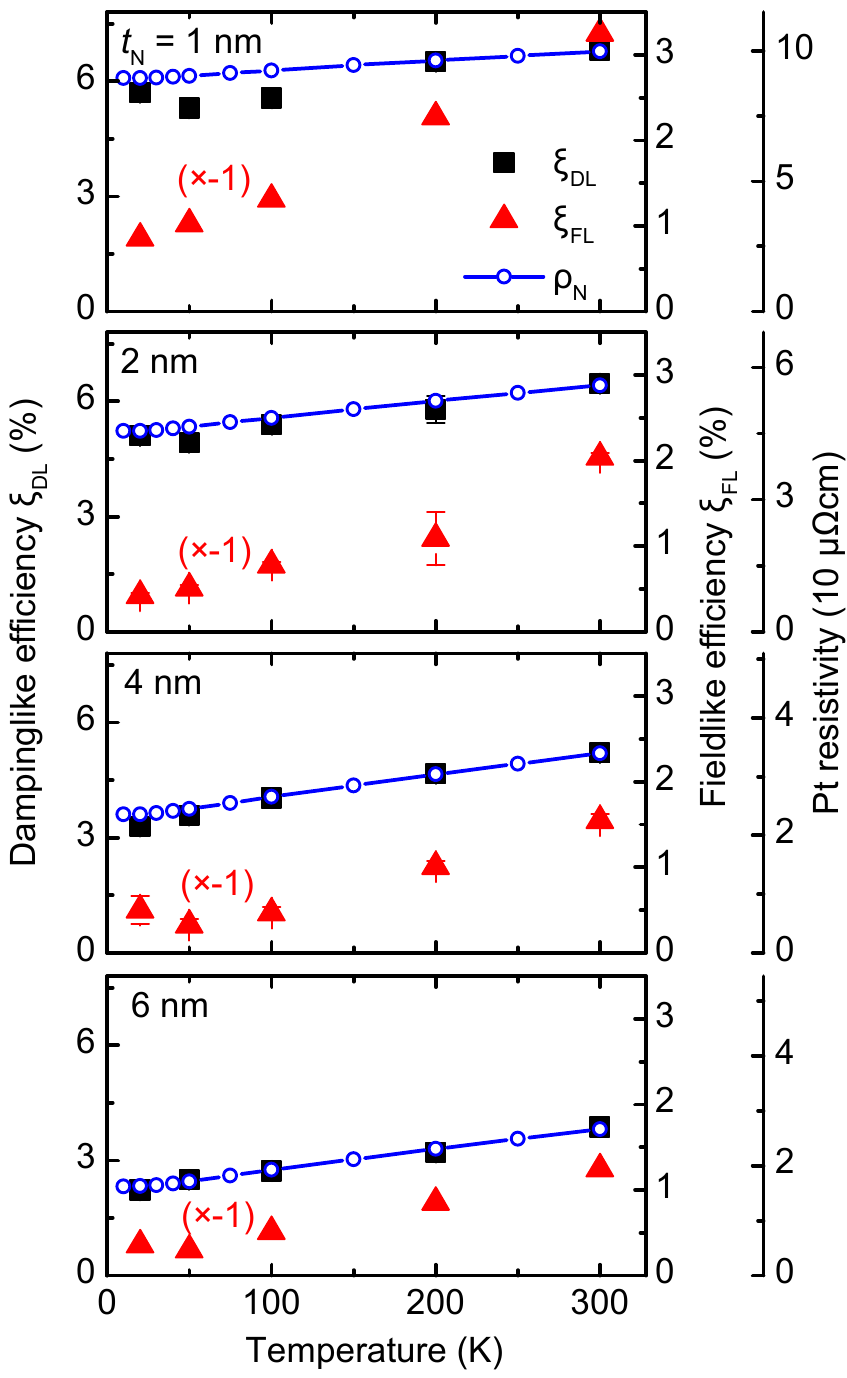}
\caption{The temperature dependence of the SOT efficiencies, $\xi_{DL}$ (black squares, left ordinate) and $\xi_{FL}$ (red triangles, right ordinate 1) for the 1, 2, 4, and 6~nm Pt bilayers as indicated on the figure. The $\xi_{FL}$ data has been scaled by a factor of $-1$. The error bars represent the standard errors. The Pt resistivity is shown (right ordinate 2) as the blue open circles, and the lines connect data points.} \label{fig:t_dep}
\end{figure}

The temperature-dependent AHE and AMR behaviors we observe are in good agreement with literature reports of a low-temperature MPE in F/Pt bilayers\cite{Huang2012, Lim2013, Miao2014, Zhang2015}, although quantitative parameters such as the magnetic moment density or MPE layer thickness are not easily extracted from these measurements. Although few experimental papers directly discuss the influence of the MPE on SOT efficiencies, \citet{Lim2013} have commented that the MPE at a F/Pt interface may affect spin-dependent transport significantly through enhanced transverse dephasing processes in the MPE Pt volume. The distinguishing experimental feature is expected to be the temperature dependence, because the MPE is enhanced at low temperatures. To study the influence of the MPE on the SOT efficiencies, we have performed the $\xi_{DL}$ and $\xi_{FL}$ harmonic response measurement from 300~K to 20~K, the results of which are summarized in Fig. \ref{fig:RT_summ}(b). Both the DL and FL components are detected for all Pt thicknesses, with $\xi_{FL}$ having opposite sign and smaller magnitude than $\xi_{DL}$. The signs\footnote{With current defining the $+x$-direction, $H_{DL}$ along the $+z$-direction gives positive $\xi_{DL}$ for magnetization along $+x$, and $H_{FL}$ along $-y$ gives negative $\xi_{FL}$. The signs of these efficiencies would be reversed if the order of the stack was reversed from F/N to N/F.} of $\xi_{DL}$ and $\xi_{FL}$ are in agreement with measurements reported for CoFe/Pt bilayers\cite{Emori_Beach_2013}. In Fig. \ref{fig:RT_summ}(b), it is clear that $\xi_{FL}$ is strongly suppressed at low temperature for all thicknesses, while $\xi_{DL}$ shows only modest suppression. The SOT efficiencies are plotted vs temperature in Fig. \ref{fig:t_dep}. 

In the discussion that follows below, we propose a mechanism by which the MPE may suppress $\xi_{FL}$ at low temperature, in which we attribute the DL SOT to the Pt SHE, and the FL SOT to the CFA/Pt interface Rashba effect. This causal distinction is well-motivated for F/Pt bilayers\cite{Liu_Pt_2012, Nguyen2016,Fan_2013,Miron2011}, and is supported by the qualitatively different trends we observe in $\xi_{DL}$ and $\xi_{FL}$ as Pt thickness and temperature are varied. In principle, the CFA/MgO interface may also possess a Rashba interaction, however as Pt thickness is increased, a diminishing fraction of the current is shunted through the CFA layer. Because the $\xi_{FL}$ data shown in Fig. \ref{fig:RT_summ}(b) plateaus for large Pt thickness \textit{when normalized by Pt current density}, the Pt and CFA/Pt interface give the dominant sources of SOTs. An alternative explanation of the FL SOT in F/N bilayers invokes the N SHE and a nonzero imaginary component of the interface mixing conductance Im$(G_{\uparrow\downarrow})$, which has been supported by recent measurements involving light-metal spacer layers\cite{Fan_2014, Nan2015, Ou2016}. We will return to a discussion of our SOT measurements in the context of the SHE-Im$(G_{\uparrow\downarrow})$ interpretation near the end of this article. 
%
%

First, we discuss the $\xi_{DL}$ measurements summarized in Fig. \ref{fig:RT_summ}(b). DL SOT efficiencies in F/N bilayers are typically interpreted through fits to the N SHE spin diffusion model\cite{Nguyen2016,Ganguly_2014}, the hallmark of which is an increase in $\xi_{DL}$ with increasing N thickness, saturating at a thickness set by the spin diffusion length. Because $\xi_{DL}$ in our samples  \textit{decreases} monotonically with increasing Pt thickness, any na\"{i}ve model would imply that a corresponding spin diffusion length is less than $\sim1$~nm. Although the data may be interpreted by invoking a spin diffusion length less than 1~nm, the value itself does not have real physical significance given that it is smaller than the momentum scattering length, which in this limit is set by the film thickness. In Fig. \ref{fig:t_dep}, the right ordinate is used to compare $\xi_{DL}$ to Pt resistivity as the temperature is varied. We see that $\xi_{DL}$ tracks $\rho_{Pt}$ closely: for small thicknesses ($t_N=1,2$~nm), where the Pt RRR is small, the temperature dependence of $\xi_{DL}$ is weak, whereas for large thicknesses  ($t_N=6,8$~nm), where the RRR is larger, $\xi_{DL}$ has a more pronounced temperature dependence. The observation that $\xi_{DL}\propto\rho$, if interpreted through the SHE diffusion model, is consistent with the intrinsic (or possibly side-jump) SHE scaling reported for Pt\cite{Nguyen_Bass_2014,Nguyen2016,Sagasta2016}. However, spin backflow could also result in a similar phenomenology, as $\xi_{DL}\propto2G_{\uparrow\downarrow}/(G_{N}+2G_{\uparrow\downarrow})$ where $G_N\equiv (\rho\lambda)^{-1}$ and $G_{\uparrow\downarrow}$ is the F/N interface spin-mixing conductance\cite{Brataas2001}. Spin backflow is significant for Pt, due to the relatively low resistivity and short spin diffusion length. From a fitting point-of-view, we cannot constrain enough parameters to distinguish between these two explanations for the $\xi_{DL}\propto\rho$ observation. Furthermore, we caution that when the SHE diffusion model parameters (SH ratio, spin diffusion length, N spin resistance) vary with N resistivity, all of the models become poorly constrained.

\begin{figure}
\includegraphics{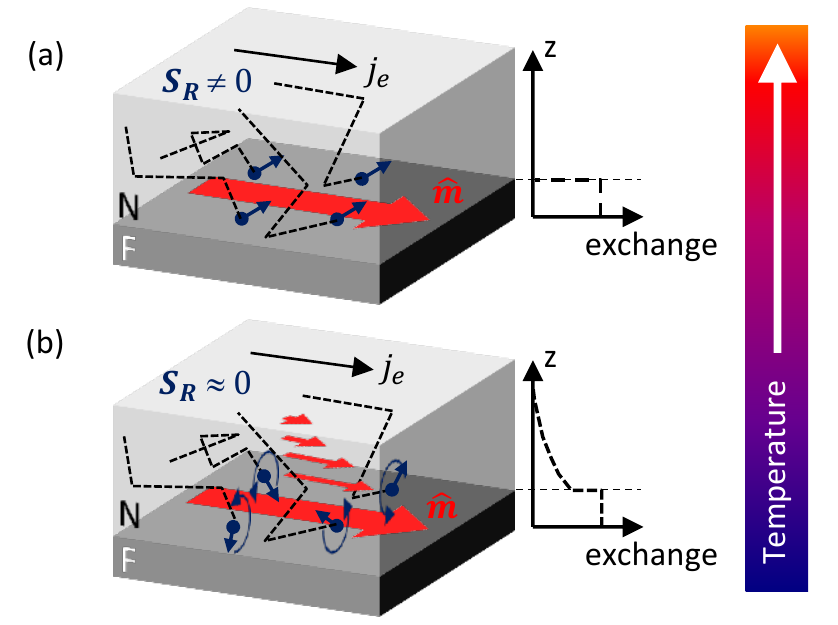}
\caption{Illustrations of (a) the Rashba spin accumulation $S_R$ at high temperature in absence of the MPE and (b) MPE order at low temperature which serves to precess and destroy the transverse Rashba spin accumulation. The magnetization $\bm{\hat{m}}$ and current $\bm{j_e}$ are indicated with red and black arrows respectively. Right-drifting carriers in N, which make up the current, are drawn as blue arrows (denoting spin accumulation) with black dashed trajectories implying scattering events. Plots of the exchange interaction strength vs. depth coordinate $z$ in the bilayer, which are schematic and not drawn to scale, are included.} \label{fig:cartoon}
\end{figure}

We now turn to discussing the temperature dependence of the FL SOT efficiency, which is shown in Fig. \ref{fig:t_dep}. For all thicknesses, the magnitude of $\xi_{FL}$ decreases by a factor nearly of $4$ from 300~K to 20~K, in contrast to $\xi_{DL}$, for which the temperature dependence simply follows the Pt resistivity. A similar behavior of $\xi_{FL}$ has been observed in annealed CoFe/Pt\cite{Pai2015}. We believe that the decrease in $\xi_{FL}$ as temperature decreases is due to the increased MPE at low temperatures. The FL component of the SOT originates from the exchange interaction between a Rashba-induced spin accumulation in N and the F magnetization\cite{Manchon2009,MihaiMiron2010}. In Fig. \ref{fig:cartoon}(a), the Rashba spin accumulation is drawn transverse to the magnetization to illustrate the maximal torque configuration in absence of the MPE. However, for nonzero MPE, the Rashba spin accumulation generated at the interface transverse to $\bm{\hat{m}}$ rapidly precesses about and is dephased by the inhomogeneous MPE exchange field, as is illustrated in Fig. \ref{fig:cartoon}(b). Perhaps counter-intuitively, at low temperatures where moments in N and F are strongly coupled, $\xi_{FL}$ decreases because the exchange interaction extends into N and destroys the spin accumulation responsible for the FL SOT. The physics of the MPE suppression of the FL SOT may not be captured by existing models, which assume an interface delta function exchange coupling between the F and N moments\cite{HaneyStiles2013,Amin2016} rather than a spatially nonuniform MPE exchange interaction extending a finite thickness into N. We note that in some cases\cite{Qiu2014, Ou2016} the FL SOT has been observed to increase with temperature in bilayers with Ta and W as the N metal, which are not believed to support MPEs. It is not clear if the FL SOTs presented in Refs. \cite{Qiu2014, Ou2016}, and their temperature-dependencies, are due to the same mechanisms as those presented in this article.

In the above discussion, we have attributed the FL SOT to the CFA/Pt interface Rashba effect. We briefly discuss the alternative picture in which the FL SOT arises from the Pt SHE through Im$(G_{\uparrow\downarrow})$.  Im$(G_{\uparrow\downarrow})$ physically corresponds to incomplete absorption of transverse spin current by the F layer, which can be pictured semiclassically as N spin current reflecting from the F layer with spin precessed through finite phase (rather than complete precessional dephasing). It is believed that Re$(G_{\uparrow\downarrow})\gg$Im$(G_{\uparrow\downarrow})$, with sizable Im$(G_{\uparrow\downarrow})$ only occurring for very thin (few-Angstrom) F layers. If we interpret our data in the picture where the FL SOT arises from the Pt SHE through Im$(G_{\uparrow\downarrow})$, the implication would be that Im$(G_{\uparrow\downarrow})$ increases as temperature is increased. The same efficient dephasing of spin accumulation transverse to $\bm{\hat{m}}$ due to the MPE can explain this trend: at low temperature, the extension of the magnetized volume into the Pt\cite{Lim2013, Klewe2015} suppresses Im$(G_{\uparrow\downarrow})$ by the increase in the effective F thickness. 

We conclude by highlighting an important distinction of the MPE precessional dephasing process from interface spin-memory loss relaxation processes\cite{Park2000}. For spin-magnetization interactions, angular momentum conservation necessitates that the MPE suppression of the transverse interface spin accumulation represents a transverse spin current sunk into the N MPE magnetization, which should result in a DL torque (as the N magnetization is exchange-coupled to the F magnetization). In the Rashba FL SOT interpretation, this would reflect a transfer of FL SOT to DL Rashba SOT, and in the SHE picture reflect a corresponding increase in Re$(G_{\uparrow\downarrow})$ as Im$(G_{\uparrow\downarrow})$ decreases. However, we observe no distinguishable increase in $\xi_{DL}$ at low temperatures. Therefore, we conclude that the MPE suppression of the Rashba spin accumulation generates a much smaller \textit{spin current} than is generated by the SHE, which is consistent with the discussion by \citet{HaneyStiles2013}. In the case of the SHE spin current generated in the bulk of N away from the interface, we expect that the few-\AA~thick MPE layer extends the effective F/N interface slightly into the Pt but does not influence $\xi_{DL}$, consistent with SOT-FMR measurements by \citet{Zhang2013} for Pt thicknesses larger than 1~nm. 

In conclusion, we have demonstrated the suppression of Rashba SOTs as the MPE increases at low temperature  in F/Pt bilayers. This identification implies engineering of the MPE may provide a technique to maximize Rashba SOT efficiencies in F/Pt bilayers.  


This work was supported by C-SPIN, one of the six centers of STARnet, a SRC
program sponsored by MARCO and DARPA, and the NSF NNCI program.
\bibliography{library}

\end{document}